\documentclass[final]{eptcs}
\providecommand{\event}{VPT 2007} 
\usepackage{stmaryrd}
\usepackage{enumitem}
\usepackage[utf8]{inputenc}
\usepackage{graphicx}
\usepackage{xspace}
\usepackage{amsmath,amsfonts,amssymb,dsfont,mathrsfs,amsthm}
\usepackage{etex}
\usepackage{xparse}
\usepackage{glossaries}
\usepackage{listings}
\usepackage{languages}
\usepackage{multicol}
\usepackage{tikz}
\usetikzlibrary{shapes,arrows}
\usetikzlibrary{matrix,arrows,decorations.pathmorphing}
\usepackage{rotate}
\usepackage{pdflscape}
\usepackage{afterpage}
\usepackage{moreverb}
\usepackage{mathtools}
\usepackage{proof}
\usepackage{setspace}
\usepackage{aliascnt}
\usepackage[numbers]{natbib}
\usepackage{multirow}
\usepackage{doi}
\usepackage{xifthen}

\setlist{noitemsep}

\DeclareMathOperator{\app}{+\!\!\!+}

\newcommand{\FramaC}{\textsc{Frama-C}\xspace}
\newcommand{\toolname}{\textsc{conc2seq}\xspace}
\newcommand{\framac}{\FramaC}

\newcommand\Coq{\textsc{Coq}\xspace}

\newcommand{\ACSL}{\textsc{acsl}\xspace}

\newcommand{\WP}{\textsc{Wp}\xspace}

\NewDocumentCommand\infrule{mmooom}{
  \IfNoValueTF{#5} 
    {\IfNoValueTF{#4}
      {\[\infer[#6{#1}]{\text{#2}} {\text{#3}}\]}
      {\[\infer[#6{#1}]{\text{#2}} {\text{#3} & \quad \text{#4}}\]}}
    {\[\infer[#6{#1}]{\text{#2}}{\deduce{\text{#3} \quad \text{#4}}{ 
\text{#5}}}\]}
}

\NewDocumentCommand\tsor{oooom}{
  \infrule{#5}{#1}[#2][#3][#4]{}
}

\NewDocumentCommand\scr{oooom}{
  \infrule{#5}{#1}[#2][#3][#4]{}
}

\NewDocumentCommand\lanr{oooom}{
  \infrule{#5}{#1}[#2][#3][#4]{}
}

\NewDocumentCommand\glor{oooom}{
  \infrule{#5}{#1}[#2][#3][#4]{}
}

\NewDocumentCommand\blor{oooom}{
  \infrule{#5}{#1}[#2][#3][#4]{}
}

\newacronym{sc}{SC}{Séquentiellement Consistant}

\newacronym{tso}{\textsc{TSO}}{Total Store Order}
 
\newacronym{pso}{\textsc{PSO}}{Partial Store Order}

\newacronym{chr}{CHR}{Constraint Handling Rules}

\NewDocumentCommand\thr{o}{
  \IfNoValueTF{#1} 
    {processus léger}
    {processus léger #1}
}

\newtheorem{theorem}{Theorem}

\newaliascnt{lemma}{theorem}
\newtheorem{lemma}[lemma]{Lemma}
\aliascntresetthe{lemma}
\newaliascnt{proposition}{theorem}

\aliascntresetthe{proposition}

\newenvironment{sproof}{%
  \proof}{\endproof}

\newcommand{\site}{http://www.univ-orleans.fr/lifo/membres/loulergu/sylvain.dailler/Modsysmem}

\newlength{\thmmargina}
\newlength{\thmmarginb}
\newlength{\thmmarginc}
\newlength{\thmmargind}
\newlength{\thmmargine}
\newlength{\thmmarginf}
\newlength{\thmmarging}
\newlength{\thmmarginh}
\newtheoremstyle{thma}{9pt}{9pt}{\itshape}{-\thmmargina}{\small\bfseries}{\quad}{ }{}
\theoremstyle{thma}

\newtheoremstyle{thmb}{9pt}{9pt}{\itshape}{-\thmmarginb}{\small\bfseries}{\quad}{ }{}
\theoremstyle{thmb}

\newtheoremstyle{thmc}{9pt}{9pt}{\itshape}{-\thmmarginc}{\small\bfseries}{\quad}{ }{}
\theoremstyle{thmc}

\newtheoremstyle{thmd}{9pt}{9pt}{\itshape}{-\thmmargind}{\small\bfseries}{\quad}{ }{}
\theoremstyle{thmd}

\newtheoremstyle{thme}{9pt}{9pt}{\itshape}{-\thmmargine}{\small\bfseries}{\quad}{ }{}
\theoremstyle{thme}

\newtheoremstyle{thmf}{9pt}{9pt}{\itshape}{-\thmmarginf}{\small\bfseries}{\quad}{ }{}
\theoremstyle{thmf}

\newtheoremstyle{thmg}{9pt}{9pt}{\itshape}{-\thmmarging}{\small\bfseries}{\quad}{ }{}
\theoremstyle{thmg}

\newtheoremstyle{thmh}{9pt}{9pt}{\itshape}{-\thmmarginh}{\small\bfseries}{\quad}{ }{}
\theoremstyle{thmh}

\newcommand{\seqname}{\mathit{seq}}
\newcommand{\parname}{\mathit{par}}
\newcommand{\simname}{\mathit{sim}}
\newcommand{\initname}{\mathit{init}}

\newcommand{\setN}{\ensuremath{\mathbb{N}}\xspace}
\newcommand{\setLoc}{\ensuremath{\mathds{L}}\xspace}
\newcommand{\setTid}{\ensuremath{\mathds{T}}\xspace}
\newcommand{\setNum}{\ensuremath{\mathbb{Z}}\xspace}
\newcommand{\setBool}{\ensuremath{\mathds{B}}\xspace}
\newcommand{\setVal}{\ensuremath{\mathcal{V}}\xspace}
\newcommand{\setVar}{\ensuremath{\mathcal{X}}\xspace}
\newcommand{\setName}{\ensuremath{\mathit{Name}}\xspace}
\newcommand{\setCode}{\ensuremath{\mathcal{C}}\xspace}

\newcommand{\setEnv}{\ensuremath{\mathcal{E}}\xspace}
\newcommand{\setCtx}{\ensuremath{\mathcal{L}}\xspace}
\newcommand{\setStack}{\ensuremath{\mathcal{S}}\xspace}
\newcommand{\setHeap}{\ensuremath{\mathcal{H}}\xspace}
\newcommand{\setStSeq}{\ensuremath{\Sigma_{\seqname}}\xspace}
\newcommand{\setStPar}{\ensuremath{\Sigma_{\parname}}\xspace}

\newcommand{\loc}{l}
\newcommand{\instr}{\mathit{instr}}

\newcommand{\vtrue}{\ensuremath{\mathbf{true}}\xspace}
\newcommand{\vfalse}{\ensuremath{\mathbf{false}}\xspace}

\newcommand{\stname}{\sigma}

\newcommand{\stsim}{\ensuremath{\stname_{\simname}}}

\newcommand{\stpar}{\ensuremath{\stname_{\parname}}}

\newcommand{\heapname}{\eta}
\newcommand{\heapseq}{\ensuremath{\heapname_{\seqname}}}
\newcommand{\heappar}{\ensuremath{\heapname_{\parname}}}
\newcommand{\heapsim}{\ensuremath{\heapname_{\simname}}}
\newcommand{\heapsimo}{\heapsim^{\parname}}
\newcommand{\heapsims}{\heapsim^{\simname}}

\newcommand{\stackname}{s}

\newcommand{\parstack}{\ensuremath{\stackname_{\parname}}}
\newcommand{\simstack}{\ensuremath{\stackname_{\simname}}}

\newcommand{\Nextlabel}{\ensuremath{\textsc{next}}\xspace}
\newcommand{\nextlabel}[1]{\Nextlabel(#1)}

\newcommand{\progname}{\mathit{prog}}
\newcommand{\sprog}{\ensuremath{\progname_{\seqname}}\xspace}
\newcommand{\pprog}{\ensuremath{\progname_{\parname}}\xspace}
\newcommand{\simprog}{\ensuremath{\progname_{\simname}}\xspace}
\newcommand{\smem}{\ensuremath{\mathit{memory}}}
\newcommand{\pmem}{\smem}

\newcommand{\emptyenv}{\ensuremath{\varnothing}\xspace}
\newcommand{\envname}{\rho}
\newcommand{\env}{\envname}

\newcommand{\Mem}{\ensuremath{\mathit{mem}}\xspace}
\newcommand{\mem}[1]{\ensuremath{\Mem_{#1}}}
\newcommand{\memf}[2]{\ensuremath{\Mem_{#1}(#2)}}

\newcommand{\Procs}{\ensuremath{\mathit{procs}}\xspace}
\newcommand{\procs}[1]{\ensuremath{\Procs_{#1}}}

\newcommand{\prog}{\ensuremath{\mathit{prog}}\xspace}
\newcommand{\Prog}{\ensuremath{\mathcal{P}}}

\newcommand{\Mains}{\ensuremath{\mathit{mains}}\xspace}
\newcommand{\MainsP}[1]{\ensuremath{\Mains_{#1}}\xspace}
\newcommand{\mains}[2]{\MainsP{#1}(#2)}

\newcommand{\Stacks}{\ensuremath{\mathit{stacks}}\xspace}
\newcommand{\StacksP}[1]{\ensuremath{\Stacks_{#1}}\xspace}
\newcommand{\stacks}[2]{\StacksP{#1}(#2)}

\newcommand{\Main}{\ensuremath{\mathit{main}}\xspace}
\newcommand{\main}[1]{\ensuremath{\Main_{#1}}\xspace}

\newcommand{\Body}{\ensuremath{\mathit{body}}\xspace}
\newcommand{\BodyP}[1]{\ensuremath{\Body_{#1}}\xspace}
\newcommand{\body}[2]{\ensuremath{\BodyP{#1}(#2)}}

\newcommand{\seqstacki}{\ensuremath{[(\Main,~\emptyenv,~\body{}{\Main})]}}
\newcommand{\thdstacki}{\ensuremath{[(\mains{}t,~\emptyenv,~\body{}{\mains{}{t}})]}}

\newcommand{\iwhile}[2]{\ensuremath{\mathbf{while}~#1~\mathbf{do}~#2}}
\newcommand{\iif}[3]{\ensuremath{\mathbf{if}~#1~\mathbf{then}~#2~\mathbf{else}~#3}}
\newcommand{\iatomic}[1]{\ensuremath{\mathbf{atomic}(#1)}}
\newcommand{\Select}{\ensuremath{\mathbf{select}}\xspace}
\newcommand{\iselect}[3]{\Select_{#1}(#2,~#3)}

\newcommand{\simmain}{\ensuremath{\mathbf{interleavings}}\xspace}

\newcommand{\numtid}{\ensuremath{\mathit{\#tid}}\xspace}

\newcommand{\asilent}{\ensuremath{\tau}}
\newcommand{\aread}[3]{\ensuremath{\mathbf{read}~#1~#2~#3}}
\newcommand{\awrite}[3]{\ensuremath{\mathbf{write}~#1~#2~#3}}
\newcommand{\acall}[2]{\ensuremath{\mathbf{call}~#1~#2}}
\newcommand{\areturn}[1]{\ensuremath{\mathbf{return}~#1}}
\newcommand{\aatomic}[1]{\ensuremath{\mathbf{atomic}~#1}}
\newcommand{\saction}{a_{\seqname}}
\newcommand{\paction}{a_{\parname}}

\newcommand{\ctxname}{\mathit{ctx}}
\newcommand{\ctxf}[3]{(#1,~#2,~#3)}
\newcommand{\ctxmr}[1]{\ctxf{m}{\envname}{#1}}
\newcommand{\ctxmrc}[1]{\ctxmr{(#1;~c)}}
\newcommand{\stackcons}[2]{#1\cdot#2}
\newcommand{\stackconss}[1]{\stackcons{#1}{s}}
\newcommand{\state}[2]{#1,~#2}

\newboolean{InArray}
\setboolean{InArray}{false}

\newenvironment{marray}[1]
{\setboolean{InArray}{true}$\begin{array}{#1}}
{\end{array}$\setboolean{InArray}{false}}

\newcommand{\ssem}[5]{#1 \vdash #2
  \ifthenelse{\boolean{InArray}}{&}{}
  \xrightarrow{~#3~}\ifthenelse{\isempty{#5}}{}{\mathrel{\vphantom{\to}^*}}
  \ifthenelse{\boolean{InArray}}{&}{}
  #4}
\newcommand{\ssemP}[3]{\ssem{\Prog}{#1}{#2}{#3}{}}
\newcommand{\psemP}[3]{\ssem{\Prog,~\numtid}{#1}{#2}{#3}{}}
\newcommand{\eevalf}[2]{\llbracket #1 \rrbracket_{#2}}
\newcommand{\eeval}[1]{\eevalf{#1}{\envname}}
\newcommand{\update}[3]{#1[#2 \mapsto #3]}

\newcommand{\lbl}{\ell}
\newcommand{\lbln}{\ell_{\mathit{next}}}
\newcommand{\linstr}{\instr^\lbl_{\lbln}}

\newcommand{\simvar}[2]{\&_{#1}#2}

\newcommand{\pct}{\ensuremath{\mathtt{pct}}\xspace}
\newcommand{\ptid}{\ensuremath{\mathtt{ptid}}\xspace}
\newcommand{\tid}{\ensuremath{\mathsf{tid}}\xspace}

\newcommand{\lblToName}{\ensuremath{\mathit{toName}}\xspace}

\newcommand{\tmp}{\ensuremath{\mathsf{tmp}}\xspace}
\newcommand{\aux}{\ensuremath{\mathsf{aux}}\xspace}
\newcommand{\terminated}{\ensuremath{\mathsf{terminated}}\xspace}
 
\newcommand{\Load}{\mathit{load}}
\newcommand{\load}[1]{\Load(#1)}

\newcommand{\From}{\ensuremath{\mathtt{from}}\xspace}
\newcommand{\from}[1]{\From(#1)}
\newcommand{\blbl}[1]{\mathit{begin}(#1)}
\newcommand{\elbl}[1]{\mathit{end}(#1)}

\newcommand{\cinit}{\ensuremath{c_{\mathit{init}}}\xspace}
\newcommand{\cselect}{\ensuremath{c_{\mathit{select}}}\xspace}
\newcommand{\csim}{\ensuremath{c_{\mathit{sim}}}\xspace}
\newcommand{\cterminate}{\ensuremath{c_{\mathit{termination}}}\xspace}

\newcommand{\Wfstack}{\textsc{wf\_stack}}
\newcommand{\wfstack}[2]{\Wfstack(#1,~#2)}

\newcommand{\trname}{\mathit{tr}}

\newcommand{\assignRule}{\ensuremath{\mathbf{[assign]}}\xspace}
\newcommand{\readRule}{\ensuremath{\mathbf{[read]}}\xspace}
\newcommand{\writeRule}{\ensuremath{\mathbf{[write]}}\xspace}
\newcommand{\selectRule}{\ensuremath{\mathbf{[select]}}\xspace}
\newcommand{\callRule}{\ensuremath{\mathbf{[call]}}\xspace}
\newcommand{\returnRule}{\ensuremath{\mathbf{[return]}}\xspace}
\newcommand{\seqRule}{\ensuremath{\mathbf{[seq]}}\xspace}
\newcommand{\atomicRule}{\ensuremath{\mathbf{[atomic]}}\xspace}

\title{%
  From Concurrent Programs to
  Simulating Sequential Programs:
  Correctness of a Transformation
}
\author{%
  Allan Blanchard
  \institute{Univ. Orl\'{e}ans,\\ INSA Centre Val de Loire\\
    LIFO EA 4022\\
    45067 Orl\'{e}ans, France}
  \and
  Fr\'{e}d\'{e}ric Loulergue
  \institute{Northern Arizona University\\
    School of Informatics\\ Computing and Cyber Systems\\
    Flagstaff, USA}
  \and
  Nikolai Kosmatov
  \institute{Software Reliability Laboratory\\
    CEA, LIST, PC 174\\
    91191 Gif-sur-Yvette, France}
}

\def\titlerunning{From Concurrent Programs to Simulating Sequential
  Programs: Correctness of a Transformation}
\def\authorrunning{A. Blanchard, F. Loulergue, N. Kosmatov}

\begin{document}

\maketitle

\begin{abstract}
  \FramaC is a software analysis framework that provides a common
infrastructure and a common behavioral specification language to
plugins that implement various static and dynamic analyses of C
programs. Most plugins do not support concurrency. We have proposed
\toolname, a \FramaC plugin based on program transformation, capable
to leverage the existing huge code base of plugins and to handle
concurrent C programs.

In this paper we formalize and sketch the proof of correctness of the
program transformation principle behind \toolname, and present an
effort towards the full mechanization of both the formalization and
proofs with the proof assistant \Coq.


\end{abstract}

\section{Introduction} 
\label{sec:introduction}
\FramaC~\cite{KPS2013:TAP,KKP2015:FAC} is a framework for static and
dynamic analysis of C programs. It offers a common infrastructure
shared by various plugins that implement specific analyses, as well as
a behavioral specification language named ACSL~\cite{ACSL:20151002}.
Developing such a platform is a difficult and time-consuming task.  As
most existing \FramaC plugins do not support {\em concurrent} C code,
extending the current platform to handle it is an interesting and
promising work direction.

Motivated by an earlier case study on deductive verification of an
operating system component~\cite{BKL2015:FMICS}, we have proposed a
new plugin, named \toolname~\cite{BKL2016:SCAM}, that allows \FramaC
to deal with concurrent programs.  In order to leverage the existing
plugins, we designed \toolname as a code transformation tool. For {\em
  sequentially consistent} programs~\cite{L79:IEEE}, a concurrent
program can be simulated by a sequential program that produces all
interleavings of its threads.

To ensure that the proofs and analyses conducted using \toolname are
correct, we need to assure that the transformation preserves the
semantics of programs. The contribution of this paper presents the
proof of correctness of the code transformation principle used in
\toolname.

The verification of the transformation is done for simplified
languages that capture the interesting property with respect to
validity, in particular memory accesses and basic data and control
structures (both sequential and parallel). We formalize the source
(parallel) language as well as the target (sequential) language and
formally define the transformation on these languages.

In these languages, we do not consider all control structures of the C
language but only simple conditionals and loops (\lstinline{goto} and
\lstinline{switch} are not part of the considered languages). The C
assignments are decomposed into three simpler constructs: local
assignments that do not incur access to the global memory, reading of
the global memory (one location at a time), and writing into the
global memory (one location at a time). An expression can only be
composed of constants, basic operations and local variables. Procedure
calls are allowed but recursion is not.  There is no dynamic memory
allocation.

In the remaining of this report, we present first the considered
source and target languages as well as their formal semantics
(Section~\ref{sec:languages}).  Then we describe the transformation
(Section~\ref{sec:transformation}).  Section~\ref{sec:correctness} is
devoted to the equivalence relation between states of the source
program and states of the transformed program, and its use for the proof of
correctness of the proposed transformation.  We discuss an ongoing
effort to mechanize the formalization and proof with the interactive
theorem prover \Coq in Section~\ref{sec:coq}.  Finally, we position
our contribution with respect to the literature in
Section~\ref{sec:relatedwork} and conclude in
Section~\ref{sec:conclusion}.


\section{Considered Languages} 
\label{sec:languages}
\subsection{Syntax and Program Definition}

We consider an enumerable set of memory locations \setLoc.  We do
not support dynamic memory allocation: the memory locations
manipulated by a program are thus known before the beginning of the
execution.  A size is associated to each allocated location, i.e. the
number of values that can be stored at this memory location. 
A location $l$ can be seen as an array in C
whose first element is $l[0]$ and whose address is $l$.

The set of values that can be assigned to variables is written \setVal
and is the union of memory locations (\setLoc), integers (\setNum) and
booleans (\setBool).  We assume that different values of the language
take the same amount of memory.

We write \setVar for the set of local variables.  In the remaining of
the paper, for a set $A$ whose elements are written $a$, $A^*$ is the
set of finite sequences of elements of $A$ and $\overline{a}$ will
denote an element of $A^*$, i.e. a sequence of elements of $A$.
Expressions are defined as follows:
$$
\begin{array}{rcll}
  \setVal \ni v & ::= & n \mid \loc \mid b
  &  n\in \setNum,~\loc \in\setLoc,~b \in \setBool\\
  e & ::= & v \mid x \mid op(\overline{e}) & x \in \setVar\\
\end{array}
$$
We do not define the set of operators here: it is a usual set of
arithmetic and boolean operations. It is however necessary to
emphasize that these operators do not allow pointer arithmetic.  The
only provided operation on memory locations is comparison.
Expressions cannot produce side-effects. In the remaining of the paper,
expressions will be denoted by $e$ and variants.

\vspace{-3mm}
\paragraph{Sequential language.} A sequential program is defined
as a sequence of procedures, by convention the first one being
the main procedure. A procedure is defined by its name, its parameters
(that are a subset of local variables) and the sequence of instructions
that form its body:
$$
\begin{array}{l@{ \quad }r@{ \quad }l r}
  \mathit{proc}  &  ::=  &  m(\overline{x}) \mathit{c}  &   m \in \setName\\
  \instr &  ::=  & x  := e  &  \text{local assignment} \\
                 &   |  &  x[y] := e  & \text{writing to the heap}\\
                 &   |  &  x := y[e]  & \text{reading from the heap} \\
                 &   |  &  \iwhile{e}{c} \quad | \quad \iif{e}{c}{c} \\
                 &   |  &  m(\overline{e})  & \text{procedure call}\\
  \setCode \ni \mathit{c} &  ::=  &  \{\} \mid \instr ;\mathit{c}   & \\
\end{array}
$$
where $\mathit{Name}$ is the set of valid procedure names. \Select,
\simmain, and names built from \setNum are all reserved
names. \setCode is the set of instruction lists, i.e. program code.

The language includes the usual primitives in a small imperative
language: sequence of instructions (we will write $\{ \instr_1; \instr_2\}$
instead of $\instr_1; \instr_2;\{\}$), conditionals, loops. Assignment is
decomposed into three distinct cases: assignment of a local variable
with the value of an expression, writing the value of an expression to
the heap, and reading a value from the heap to a local variable.
Expressions cannot contain reads from memory, nor procedure calls.  A
C assignment containing several accesses to the heap should therefore
be decomposed into several reads into local variables and an
assignment of an expression to a local variable, and finally, if
necessary, a write to the heap from a local variable.  Procedures can
be called using the classical syntax $m(\overline{e})$ where
$\overline{e}$ is the list of expressions passed in
arguments. Arguments are passed by value.

A sequential program \sprog is fully defined by:
\begin{itemize}[itemsep=0pt]
\item the list of its procedures (the main one taking no parameter),
\item a list of allocated memory locations with their associated sizes
  (positive numbers).
\end{itemize}
$$
\begin{array}{rcl}
  \smem & ::= & [(\loc_1,size_{\loc_1}); \dots ;(\loc_m,size_{\loc_m})]\\
  \sprog   & ::= & \overline{\mathit{proc}}\ \smem\\
\end{array}
$$

\vspace{-3mm}
\paragraph{Parallel language.} A parallel program can be executed by
any strictly positive number of threads. There is no dynamic creation
of threads. During the execution of a parallel program the number of
threads remains constant, given by a specific parameter of each execution.
Let \numtid denote this static number of threads.

$\setTid$ is the set of thread identifiers. We identify $\setTid$ with
\setN seen as subset of \setNum.  An element of $\setTid$ is thus a
{\em value} for both languages.  A parallel program can use any of the
sequential program constructs. In addition, it can contain the
instruction $\iatomic{c}$ that allows to run a sequence of
instructions $c$ atomically. In such a code section, no thread, other
than the one that initiated the execution of the atomic block, can be
executed.

A parallel program \pprog is fully defined by:
\begin{itemize}[itemsep=0pt]
\item the list of its procedures,
\item a list of allocated memory locations in the shared memory with
  their associated sizes,
\item a mapping from thread identifiers to defined procedure names,
  defining the main procedure of each thread.
\end{itemize}
$$
\begin{array}{rclr}
  \pprog & ::= & \overline{\mathit{proc}}\ \pmem\ \Mains
         & \qquad \text{ where } 
  \Mains : \setTid \rightarrow \setName
\end{array}
$$
For a program \prog (either sequential or parallel), \mem{\prog}
denotes the allocated memory of the program. This association list is
also considered as a function, therefore $\memf{\prog}{\loc}$ denotes
the size allocated for memory location $\loc$, if defined.
$\procs{\prog}$ denotes the sequence of procedures of the program.  For a parallel
program \MainsP{\prog} is the mapping from \setTid to \setName, and
for a sequential program \main{\prog} is the main procedure name.  For
a name $m$ and a program \prog, $\body{\prog}{m}$ denotes the body of
the procedure named $m$ in the program \prog.
If it is clear from the context \prog may be omitted.

\vspace{-3mm}
\paragraph{Comparison with the concurrent C of the \FramaC plugin.}
For sequential programs, the simplifications with respect to the
subset of C handled by our \toolname plugin are essentially that we do
not support pointer arithmetic, the expressions containing several
memory reads or procedure calls should be decomposed, and we
support only the ``most structured'' control structures.  The typing is
also very basic: variables and heap locations accept any type of values
(integers, booleans, memory locations) and the type of expressions is
checked dynamically by the semantic rules if necessary (for example
the expression that is a condition of a loop or conditional should
evaluate to a boolean value).

In C11, sequentially consistent concurrent atomic operations are often
described by an equivalent sequential C program that is supposed to be
atomically executed. In our \FramaC plugin, such operations are
specified using ACSL and their calls placed into atomic sections. In the
small imperative parallel language presented above, we could use the
same technique: implement atomic operations as their sequential
counterparts and put their calls into atomic blocks. For example, we
illustrate the atomic transfer of the value of an global variable to
another one in Figure~\ref{fig:atomic_exchange}. It is composed of two
instructions that are executed in a single atomic step. The resulting
simulating code will be commented later.

\begin{figure}
  \begin{minipage}{.40\textwidth}
    \begin{C2S}
atomic_transfer(l1, l2){
  @$\ensuremath{\mathbf{atomic}}($@v1 := l1[0];
        l2[0] := v1;@$)$@
}\end{C2S}
  \end{minipage}
  ~~~
  \begin{minipage}{.55\textwidth}
    \begin{C2S}
sim_atomic_2(@$\tid$@){
  @$\tmp$@ := @$\simvar{}{\texttt{l1}}$@; l1  := @$\tmp$@[@$\tid$@];
  v1 := l1[0];
  @$\tmp$@ := @$\simvar{}{\texttt{v1}}$@; @$\tmp$@[@$\tid$@] := v1;
  @$\tmp$@ := @$\simvar{}{\texttt{l2}}$@; l2  := @$\tmp$@[@$\tid$@];
  @$\tmp$@ := @$\simvar{}{\texttt{v1}}$@; v1  := @$\tmp$@[@$\tid$@];
  l2[0] := v1;
  @$\tmp$@ := @$\pct$@; @$\tmp$@[@$\tid$@] := 4;
}\end{C2S}
  \end{minipage}
  \caption{
    Atomic transfer of the value of a global variable to another, and
    the resulting simulating method for the corresponding atomic instruction.
  }
  \label{fig:atomic_exchange}
\end{figure}

In our case studies, the concurrent C programs do not need to know the
number of threads, and actually do not depend on the number of threads
except for one specific feature: global variables that are \textit{thread
local}. This kind of variables are in shared memory, but each thread
has its own independent copy. This is particularly useful to have
thread dedicated copies of global variables such as
\lstinline[language=C]"errno". In this case, in our memory model it
would mean that the number of memory locations called
\lstinline[language=C]"errno" would be dependent on the number of
threads.
The set of allocated memory locations does not depend on the
number of threads.

If we want to
model a procedure \lstinline[language=C]"f" that uses a thread local
variable \lstinline[language=C]"tlv" we can define in our parallel
language a procedure $f$ that takes an additional argument
$\mathit{tlv}$ and use, for each thread, a different main procedure
calling $f$ with a specific allocated memory location passed to
argument $\mathit{tlv}$.

However the set of allocated memory locations (as well as the number
of different main procedures) is not dependent on the number of
running threads. We can then imagine to have a kind of extended
parallel language which could contain symbolic names for thread local
variables and a pre-processor that, for a specific value of \numtid,
would generate programs of the proposed parallel language (generating
as many memory locations and main procedures as necessary).  As the
transformation presented in Section~\ref{sec:transformation} from the
proposed parallel language to the proposed sequential language also
depends on \numtid, we do not consider this aspect to be a limitation
of our modelling approach. These modelling choices allow to keep both
languages simple and representative. 

\subsection{Semantics}


For a sequential program, or a thread, the local environment
$\envname$ is a partial function from local variables to values:
$\envname: \setVar \rightharpoonup \setVal$. The set of local
environments is written \setEnv. \emptyenv denotes the empty
environment, i.e. the function undefined everywhere.

For both the sequential and the parallel languages, a heap
$\heapname: \setLoc \rightharpoonup \setN \rightharpoonup \setVal$ is a
partial function from memory locations that returns a partial function
from indices to values, thus essentially defining an array indexed
from $0$. \setHeap is the set of heaps.  For a defined memory
location, the associated partial function is defined continuously for
indices from $0$ to a fixed size.

A {\em local execution context} is composed of the name of the
procedure being executed, a local environment and the code that
remains to execute. The set of local execution contexts is
$\setCtx = \setName \times \setEnv \times \setCode$.  A call stack is
defined as a sequence (stack) of local execution contexts:
$s \in \setStack = \setCtx^*$.

The states of sequential and parallel programs are respectively:
$$
\begin{array}{ccc}
\begin{array}{rcl}
  \setStSeq  & = & \setStack \times \setHeap
\end{array} & \qquad &
\begin{array}{rcl}
  \setStPar  & = & (\setTid \rightharpoonup \setStack) \times \setHeap
\end{array}
\end{array}           
$$

For a parallel state $\stpar\in\setStPar$, we denote by \StacksP{\stpar}
the first component of the state, i.e.  the mapping from thread
identifiers to stacks of local execution contexts.  We omit the index \stpar{}
when it is clear from the context.

\vspace{-3mm}
\paragraph{Initial contexts and states.}
The initial execution stack is $\seqstacki$ for a sequential program.
For a parallel program, the initial context of a thread $t\in\setTid$
is \thdstacki.  For a sequential program, an initial state is thus:
$(\state{\seqstacki}{\heapseq^{\initname}})$.
For a parallel program, an initial state is
$(\state{\StacksP{\initname}}{\heappar^{\initname}})$ where
$\forall t \in \setTid.~\stacks{\initname}{t} = \thdstacki$.

An initial heap $\heapseq^{\initname}$ should satisfy the memory
allocation defined by a sequential program, i.e. if
$(l, \mathit{size}) \in mem$ then $\heapseq^{\initname}(l)(i)$ is
defined for all $0 \le i<\mathit{size}$. In addition, the values contained
in such a memory location cannot be themselves memory locations (but
they can be any other values). The same constraints hold for an initial
heap of a parallel program.

\vspace{-3mm}
\paragraph{Final states and safe execution}
The final state of a sequential program is such that
$\exists \heapname.~\stname_{\seqname}^{\mathit{final}} =
(\state{[]}{\heapname})$ and the final state of a parallel program is
such that
$\exists \heapname.~\stname_{\parname}^{\mathit{final}} =
(\state{\Stacks}{\heapname})$ with
$\forall t \in \setTid.~\stacks{}{t} = []$.

We define a blocking state as a non final state reached from an
initial state such that no semantic rule can make the execution
progress. A {\em safe program} is a program that does not reach a
blocking state from any initial state. In particular, a safe program 
can have non-terminating executions.

\vspace{-3mm}
\paragraph{Actions}
The sequential programs produce 5 basic actions: silent action,
procedure call, procedure return, memory reading, memory writing.  For
parallel programs, the atomic block structure requires to have an
action list as a possible action:
$$
\begin{array}{lcr}
  \begin{array}{rcl}
    \saction
    & ::=
    & \asilent
      \mid \acall{m}{\overline{v}} 
      \mid \areturn{m}
      \mid \aread{\loc}{n}{v} 
      \mid \awrite{\loc}{n}{v}
  \end{array}
  & \qquad
  &\begin{array}{rcl}
     \paction
     & ::=
     & \saction \mid \aatomic{\overline{\saction}}
   \end{array}
\end{array}
$$
Execution traces are action lists for sequential programs and lists of
events, i.e. pairs of thread identifier and action, for parallel
programs.

\vspace{-3mm}
\paragraph{Operational semantics}
The operational semantics of sequential programs is defined in
Figure~\ref{fig:preuve_transfo::langage::sem_seq} (rules for loops and
conditionals are omitted, see~\cite{BKL2017:RR}).  A judgement of the
sequential semantics has the following form:
$\ssemP{\state{s}{\heapname}}{\saction}{\state{s'}{\heapname'}}$, meaning that a
new state $(\state{s'}{\heapname'})$ is reached from the state
$(\state{s}\heapname)$ and this execution step produces an action
$\saction$.  $\Prog$ is a program definition. We write
$\ssem{\Prog}{\state{s}{\heapname}}{\overline{\saction}}{\state{s'}{\heapname'}}{1}$
for the reflexive and transitive closure of the relation defined by
the inference system of
Figure~\ref{fig:preuve_transfo::langage::sem_seq}.

\begin{figure}[t]
  \centering
  \resizebox{\textwidth}{!}{
    \begin{marray}{l c r}
      \ssemP
      {\state{\stackconss{\ctxmrc{x:=e}}}{\eta}}
      {~~~~\asilent~~~~}
      {\state{\stackconss{\ctxf{m}{\update{\rho}{x}{v}}{c}}}{\eta}}
      \\
      \assignRule
      & \multicolumn{2}{l}{\text{if }\eeval{e} = v}
      \\[5mm]
      \ssemP
      {\state{\stackconss{\ctxmrc{x[e_o]:=e_v}}}{\eta}}
      {\awrite{\loc}{o}{v}}
      {\state{\stackconss{\ctxmr{c}}}{\update{\eta}{(l,o)}{v}}}
      \\
      \writeRule  &
      \multicolumn{2}{l}{\text{if }
        \eeval{e_v} = v,~
        \eeval{e_o} = o,~
        \rho(x) = \loc,~
        o < \memf{}{\loc}}
      \\[5mm]
      \ssemP
      {\state{\stackconss{\ctxmrc{x:=y[e_o]}}}{\eta}}
      {\aread{\loc}{o}{v}}
      {\state{\stackconss{\ctxf{m}{\update{\rho}{x}{v}}{c}}}{\eta}}
      \\
      \readRule &
      \multicolumn{2}{l}{
        \text{if }
        \eeval{e_o} = o,~
        \rho(y) = \loc,~
        o < \memf{}{\loc},~
        \eta(\loc)(o) = v
      }
      \\[5mm]
      \ssemP
      {\state{\stackconss{\ctxmrc{m'(\overline{e})}}}{\eta}}
      {\acall{m'}{\overline{v}}}
      {\state{
          \stackcons
          {\ctxf{m'}{\update{}{\overline{x}}{\overline{v}}}{c_{m'}}}
          {\stackconss{\ctxmr{c}}}
        }{\eta}}
      \\
      \callRule  &
      \multicolumn{2}{l}{%
        \text{if }m'(\overline{x})c_{m'} \in \Prog,~
        |\overline{x}| = |\overline{e}|,~
        \overline{\eeval{e}} = \overline{v},~
        m'\not\in s
      }
      \\[5mm]
      \ssemP
      {\state{\stackconss{\ctxmr{[]}}}{\eta}}
      {\areturn{m}}
      {\state{s}{\eta}}
      \\
      \returnRule
      \\[5mm]
      \ssemP
      {\state{\stackconss{\ctxmrc{\iselect{\numtid}{\mathit{tid}}{\mathit{pc}}}}}{\eta}}
      {\acall{\mathbf{select}}{[\loc_{\mathit{tid}},~\loc_{\mathit{pc}}]}}
      {\state{\stackconss{\ctxmr{c}}}{\update\eta{(\loc_{\mathit{tid}},0)}{t}}}
      \\
      \selectRule &
      \multicolumn{2}{l}{%
        \text{if }
        \eeval{\mathit{tid}} = \loc_{\mathit{tid}},~
        \eeval{\mathit{pc}} = \loc_{\mathit{pc}},~
        0\leq t < \numtid,~
        \eta(\loc_{\mathit{pc}})(t) \not= 0
      }
    \end{marray}}
  \caption{Operational semantics of sequential programs}
  \label{fig:preuve_transfo::langage::sem_seq}
\end{figure}

We use the following notations: $l_1 \app l_2$ is the concatenation of
two sequences/lists.  To add an element on top (i.e. on the left) of a sequence, we use
the separator ``$;$'' for sequences of instructions, and the separator ``$\cdot$''
for sequences of local contexts (stacks). $|l|$ is the length of the
sequence $l$. We write $x\in l$ to denote that $x$ is an element of
the sequence $l$, and by abuse of notation, that $x$ is a
component of a tuple in the list of tuples $l$.
$\update{f}{a}{b}$ is the function $f'$ such that  $f'(a) = b$ and
for all elements $a'$ different from $a$, we have $f'(a') = f(a')$.  For two
sequences $\overline{a}$ and $\overline{b}$ of equal length, we write
$\update{f}{\overline{a}}{\overline{b}}$ instead of
$\update{\update{f}{a_1}{b_1}\ldots}{a_n}{n_n}$.  Thus
$\update{\envname}{x}{v}$ denotes an update of variable $x$ with value $v$
in environment $\envname$ while $\update{\heapname}{(l,o)}{v}$ denotes an update
at offset $o$ of memory location $l$ with value $v$ in heap $\heapname$.
When it is the empty environment that is updated, we omit it.

$\eeval{e}$ corresponds to the evaluation of expression $e$ in local
environment $\envname$. We omit the definition of this evaluation that is
classic. For example for a variable $x$, $\eeval{x} = \envname(x)$.

This semantics is rather usual, but condition $m'\not\in s$ in rule
$\mathbf{[call]}$ forbids recursive procedure calls. Moreover there is
a special procedure call:
$\iselect{\numtid}{\mathit{tid}}{\mathit{pc}}$.  This is the only
non-deterministic rule of the sequential language.  It selects
randomly a value $t$ between $0$ and \numtid (excluded), such that
$\mathit{pc}$ is a memory location which is defined at index $t$ and
contains a value different from $0$ (reserved for terminated threads).
The memory location $\mathit{tid}$ is updated with this value $t$.
This procedure call will be used in the simulation to model the change
of current thread. Note that this procedure is not supposed to be called
in parallel programs.

\begin{figure}[h]
  \centering
  {\begin{marray}{lcr}
      \psemP
      {\state{\Stacks}{\eta}}
      {~~~~(t,\saction)~~~~}
      {\state{\update{\Stacks}{t}{s'}}{\eta'}}
      \\
      \seqRule & 
      \multicolumn{2}{l}{%
        \text{if }
        \setboolean{InArray}{false}
        \ssemP
        {\state{\stacks{}{t}}{\eta}}
        {~\saction~}
        {\state{s'}{\eta'}}
        \setboolean{InArray}{true}
        \text{ and } 0\leq t < \numtid
      }
      \\[6mm]
      \psemP
      {\state{\Stacks}{\eta}}
      {~(t,\aatomic{\overline{\saction}})~}
      {\state{\update{\Stacks}{t}{\stackconss{\ctxf{m}{\rho'}{c}}}}{\eta'}}
      \\[1mm]
      \atomicRule & 
      \multicolumn{2}{l}{%
        \text{if }
        \setboolean{InArray}{false}
        \ssem
        {\Prog}
        {\state{[\ctxmr{c_{\mathit{atomic}}}]}{\eta}}
        {~~\overline{\saction}~~}
        {\state{[\ctxf{m}{\rho'}{[]}]}{\eta'}}
        {closure}
        \setboolean{InArray}{true}
      }
      \\[1mm] &
      \multicolumn{2}{l}{%
        \text{where }
        \stacks{}{t} =
        \stackconss{\ctxmrc{\iatomic{c_{\mathit{atomic}}}}}
        \text{ and } 0\leq t < \numtid
      }
    \end{marray}
  }
  \caption{Operational semantics of parallel programs}
  \label{fig:preuve_transfo::langage::sem_par}
\end{figure}

Figure~\ref{fig:preuve_transfo::langage::sem_par} presents the
semantics of parallel programs.  A judgement of this semantics have
the following form:
$\psemP{\state{\Stacks}{\heapname}}{(t,\paction)}{\state{\Stacks'}{\heapname'}}$,
where we recall that \numtid is a strictly positive number of threads.

A thread $t$ is selected such that $0\leq t<\numtid$ and $t$ has code
to execute.  If the first instruction of $t$ is not an atomic block,
then the state is reduced using the semantics of the sequential
language. In this case the whole shared heap is given as the heap of
the sequential reduction.  The action of the sequential reduction is
combined to the thread identifier $t$ to form the event of the
parallel reduction.

If the first instruction of $t$ is an atomic block, then we use the
sequential semantics to reduce the whole block. As we reduce the whole
instruction sequence without allowing for a change of thread, the
execution of this sequence is indeed atomic. The nesting of atomic
blocks is not allowed: our semantics would be stuck in this case.


\section{Program Transformation} 
\label{sec:transformation}
Let us consider a parallel program $\overline{\Procs}~\pmem~\Mains$.
The memory of the simulating sequential program contains: \pmem, a
fresh memory location \pct of size \numtid, a fresh memory location
\ptid of size $1$, for each procedure $m$ a fresh memory location
$\from{m}$ of size \numtid (with $\From : \setName \rightarrow \setLoc$).
$\pmem$ will be shared by the threads. The
array \pct contains for each thread identifier $t$ (therefore at index
$t$) the simulation of the program counter of the thread identified by
$t$, while \ptid contains the identifier of the current running
thread. $\from{m}$ is used to manage the return of calls to $m$ in the
simulating code.

The three assignment instructions are supposed to be atomic. For loops and
conditionals, the evaluation of the condition is supposed
to be atomic.  The transformation essentially translates each {\em
  atomic instruction} of each procedure of the parallel program into
one {\em procedure} of the simulating sequential program.  This
procedure has a parameter \tid that is supposed to be the identifier
of the active thread running the instruction.  In the remaining of the
paper, variables written is $\mathsf{sans}$-$\mathsf{serif\ font}$ are fresh variables
not used in the input parallel program, but that we need to implement in
the simulating sequential program, such as \tid.

We assume that the input parallel program is labeled: each instruction
$\instr$ is labeled by two values of $\setNum\setminus\{0\}$ ($0$ is a
label that indicated termination), such that the first one, denoted
$\lbl$, is a unique label in the program definition, and the second
one, denoted $\lbln$, is the label of the instruction that follows the
current instruction in the program text (for example the label of the
next instruction of a conditional is the instruction that follows the
conditional, not the label of one of the branches).  We write
$\linstr$ for such a labeled instruction.  One important point is that
the label $\lbln$ of the last instruction of {\em each} procedure is a
label distinct from all the labels in the program.  $\blbl{m}$ is a
function that returns the label of the first instruction of the body
of procedure $m$.  $\elbl{m}$ returns the label $\lbln$ of the last
instruction of the procedure body.  If the body is empty, both
functions returns a label distinct from all other labels in the
program.

For each local variable $x$ of the program (uniquely identified by the
name $m$ of the procedure in which it appears and its name $x$),
including procedure formal parameters, we need a fresh memory location
$\simvar{m}{x}$ of allocated size \numtid (we omit $m$ in the
remaining of the paper), so that each simulated thread has a copy of
what was a local variable in the parallel program.

We detail how the transformation proceeds on an example instruction:
$(x := y + 1)^\lbl_{\lbln}$.  This instruction will be transformed
into a procedure named $\lbl$ with parameter \tid (we assume a
coercion \lblToName from \setNum to \setName, and we omit it most of
the time). $y$ is simulated by the array $\simvar{}{y}$. As reads from
the heap are not allowed in expressions, in the simulated code we
first need to read the value from $\simvar{}{y}$.  We write this
sequence of instructions $\load{y}$ defined as
$\tmp := \simvar{}{y};~y:=\tmp[\tid]$.  Note that after this sequence
of instructions, variable $y$ is defined, therefore the original
expression can be used as is.  The original assignment however should
be translated too as $x$ is simulated by an array $\simvar{}{x}$. We
translate it to: $\tmp := \simvar{}{x};~\tmp[\tid] := y+1$.  Finally
we update the program counter of the running thread, so the full
translation of the instruction is:
$$\lbl(\tid) \{~%
\tmp := \simvar{}{y};~y:=\tmp[\tid];~\tmp := \simvar{}{x};~\tmp[\tid]
:= y+1;~%
\tmp := \pct;~\tmp[\tid] := \lbln~\}$$ The generalization to an
arbitrary $x:=e$ is just that we ``load'' all the variables of $e$
before using $e$. Reading from the heap and writing to the heap are
translated in a very similar way. Figure~\ref{fig:atomic_exchange}
provides a more complex example with the simulating code of the atomic
memory transfer. 

Both conditional and loops are translated into a procedure that
evaluates the condition and then updates the program counter to the
appropriate label. For example, if the condition of a conditional is
true then the program counter is updated to the label of the first
instruction of the ``then'' branch of the original conditional, if
this branch is non-empty, otherwise the label used is the label of the
instruction that follows the original conditional.

Each procedure call is translated into one procedure that passes the
values to parameters and updates the program counter to the first
instruction of the body original procedure (label $\blbl{m}$ for a
call to $m$). Also for each procedure $m$ we generate an additional
procedure, named $\elbl{m}$, that manages the return of calls to $m$.
This procedure should be able to update the program counter to the
instruction that follows the call. To be able to do so for any call,
this return procedure should use a label previously stored at memory
location $\from{m}$ by the generated procedure that prepares the call:
$$\elbl{m}(\tid)\{~\tmp:=\from{m};~\aux:=\tmp[\tid];~\tmp:=\pct;~\tmp[\tid]:=\aux~\}$$

One procedure is generated for each atomic block. Each instruction in
the block is generated in a similar way as previously described but no
update to the program counter is done, conditionals and loops keep
their structure and their blocks are recursively translated in the
atomic fashion. Procedure calls are {\em inlined} and the body of the
called procedure is translated in the atomic fashion.  It is necessary
that procedures are not recursive for this inlining transformation to
terminate.

\begin{figure}
  \begin{C2S}
@$\simmain()$@{
  // @$\cinit$@
  @$\tmp$@:=@$\pct$@;
  @$\forall t \in [0,\numtid[,~\tmp[t]$@:=@$\blbl{\mains{}{t}}$@;
  @$\forall t \in [0,\numtid[,~\tmp$@:=@$\from{\mains{}{t}}$@; @$\tmp[t]$@ := 0;
  @$\terminated$@:=@$\vfalse$@;
  while @$\neg\terminated$@ do {
    // @$\cselect$@
    @$\iselect{\numtid}{\ptid}{\pct}$@;                                     
    // @$\csim$@
    @$\tmp$@:=@$\ptid$@; @$\tid$@:=@$\tmp[0]$@; @$\tmp$@:=@$\pct$@; @$\aux$@:=@$\tmp[\tid]$@;
    switch @$\aux$@ is { @$\lbl: \lblToName(\lbl)(\tid)$@ }
    // @$\cterminate$@
    @$\terminated$@:=@$\vtrue$@; @$\tmp$@:=0;
    while @$\tmp < \numtid$@ do {
      if @$\pct[\tmp]\not=$@0 then { @$\terminated$@:=@$\vfalse$@ } else { };
      @$\tmp$@:=@$\tmp$@+1;
    }
  }
}
\end{C2S}
  \caption{Main procedure of the simulating sequential program}
  \label{fig:simmain}
\end{figure}

Finally the main procedure of the simulating sequential program, named
\simmain, is generated (Figure~\ref{fig:simmain}). It has basically
two parts: in the first part (denoted by \cinit) each program counter
is updated to the identifier of the first instruction of the main
procedure of the considered thread. \cinit places the value at
location $\from{\mains{}{t}}$ to $0$ to stop the execution when the
main procedure ends.  \cinit also initializes the local variable
\terminated, that indicates if all threads are terminated, to \vfalse.
We suppose that there is at least one thread with a main procedure to
execute. If it were not the case, we would initialize it to \vtrue.
The second part is the main simulating loop: if there are still
threads to run, a thread identifier of an active thread is chosen
(call to \Select, instruction named \cselect), then the value $\lbl$
of the program counter for this thread is read and a switch
(it is implemented as nested conditionals, we use it here for the ease
of presentation) calls the appropriate procedure
named $\lbl$ (sequence of instructions named \csim).  The body of this
loop ends by updating the flag that indicates if there are still
running threads (sequence of instructions named \cterminate).


\section{Correctness} 
\label{sec:correctness}

To state the correctness theorem, we need two notions of equivalence:
state equivalence, relating states of the input parallel program and
states of the simulating sequential program, and trace equivalence
that relates traces generated by the input parallel program to traces
generated by the simulating sequential program. 

\subsection{Equivalence of states and traces}

We note $\stsim$ the sequential program state
$(\state{\simstack}{\heapsim})$ of the simulation of a safe parallel
program in a state $\stpar = (\state{\parstack}{\heappar})$.  In
$\heapsim$, we distinguish two disjoint parts $\heapsimo$ that
replicates $\heapname_{\parname}$ and $\heapsims$ the addresses that
simulate the local variables of $\parstack$. This second part also
includes $\pct$, $\ptid$, and the addresses $\from{m}$.  The syntax
$\heapsims[t]$ allows to partially apply $\heapsims$ for the thread
$t$ to select the part that simulates $t$. So the function
$\heapsims[t](l)$ is $\heapsim(l,t)$. We define state equivalence as
follows:

\begin{align}
  & \heapname_{\parname} = \heapsimo
    \label{eq:equiv:heap}
  \\[1mm]
  & \forall t \in \setTid,~\env \in \stacks{}{t},~x \in \setVar.~
    \rho(x) = v \Longrightarrow \heapsims[t](\&x) = v
                                                \label{eq:equiv:vars}
    \\[1mm]
  & \forall t \in \setTid,~\ctxname \in \setCtx,~\stackname\in\setStack.~
            \stacks{}{t} = \ctxname \cdot \stackname
            \Longleftrightarrow
    \heapsims[t](\pct) = \nextlabel{\ctxname}
    \tag{3a}
    \label{eq:equiv:pct:running}
    \\[1mm]
  & \forall t \in \setTid.~
    \stacks{}{t} = [] \Longleftrightarrow \heapsims[t](\pct) = 0
    \tag{3b}
    \label{eq:equiv:pct:terminated}
    \refstepcounter{equation}
    \\[1mm]
  & \forall t \in \setTid.~
    \wfstack{\stacks{}{t}}{\heapsims[t]}
    \label{eq:equiv:wfstacks}
    \\[1mm]
  & \simstack =
    (\simmain,~\env_{\simname},
    ~\{\iwhile{\neg\terminated}{(\cselect\app\csim\app\cterminate)}\})\ \wedge
    \label{eq:equiv:simloop}
    \\
  & \qquad \ 
    (\env_{\simname}(\terminated) = \vtrue
    \Longleftrightarrow
    \forall t \in \setTid.~ \heapsims[t](\pct) = 0)
    \notag
  \\[1mm]
  \hline
  & \notag \hspace*{.375\textwidth} \stpar \sim \stsim 
\end{align}
\eqref{eq:equiv:heap} expresses the fact that the original heap should
be a sub-part of the simulating heap.  For each thread $t$,
\eqref{eq:equiv:vars} relates the content of every local variable $x$
of $t$ by the content of the global array $\simvar{}{x}$ in
$\heapsims[t]$ that simulates it.


Program counters must be correctly modeled,
\eqref{eq:equiv:pct:running} and \eqref{eq:equiv:pct:terminated}
express that each program counter must point on the $\Nextlabel$
instruction to execute by thread $t$ if any
\eqref{eq:equiv:pct:running}, $0$ if not
\eqref{eq:equiv:pct:terminated}. Call stacks must be correctly modeled
by $\from{m}$ \eqref{eq:equiv:wfstacks}.  We refer
to~\cite{BKL2017:RR} for the formal definitions of $\Nextlabel$, that
returns the label of the next instruction to execute in a non-empty
local execution context, and $\Wfstack$, that relates the call stacks
of the parallel state with the labels at memory locations $\from{m}$.
Finally in condition~\eqref{eq:equiv:simloop}, the equivalence is
defined for simulating program states such that the next step to
perform is the evaluation of the condition of the loop since the
simulation of an instruction is the execution of this evaluation
followed by the body of the loop.

The equivalence of traces is defined on filtered lists of actions
generated by the semantics. In the simulating program executions, we
ignore $\asilent$-actions and memory operations in $\heapsims$. We
ignore all call to and return from simulating procedures except for
calls to \Select, and procedures that simulate the start of a call and
the return of a call.


\subsection{Correctness of the simulation}

\begin{theorem}[Correct simulation]
  \label{thm:simulation}
  Let \pprog be a safe parallel program, $\simprog$ its simulating
  program, $\stname_{\parname}^{init}$
  (resp. $\stname_{\simname}^{init}$) an initial state of
  $\pprog$ (resp. $\simprog$).

  \begin{enumerate}[label=\roman*.]
  \item From $\stname_{\simname}^{init}$, we can reach, by the
    initialization sequence \cinit, $\stsim^0$ equivalent to
    $\stpar^{\initname}$. \label{item:init}
  \item For all $\stpar$ reachable from $\stpar^{\initname}$, there
    exists an equivalent $\stsim$ reachable from $\stsim^0$ with an
    equivalent trace (Forward simulation). \label{item:forward}
  \item For all $\stsim$ reachable from $\stsim^0$, there exists an
    equivalent $\stpar$ reachable from $\stpar^{\initname}$ with an
    equivalent trace (Backward simulation). \label{item:backward}
  \end{enumerate}
\end{theorem}

The proof of this theorem rely on two main observations. First, the
parallel semantics is deterministic except for the choice of the
thread, which is not an operation of the program. Equivalently, the
only non-deterministic operation of the simulation is the call to
\Select, that models the non-deterministic behavior of the parallel
semantics. Second, once the parallel semantics has selected a thread,
the reduction is delegated to the sequential semantics that is
deterministic. The corresponding simulating code, that solves the
program counter and execute the simulating procedure, is also
deterministic. Now, if we prove the forward simulation for a
transformation and the resulting code is deterministic, then we also
prove the backward simulation, as pointed by~\cite[Def 5.]{LER2009:JAR}.
More detailed proofs of theorem~\ref{thm:simulation} can be found
in the report~\cite{BKL2017:RR}.

We show that the initialization establish the
equivalence~\eqref{item:init} by induction on traces.  For the forward
simulation~\eqref{item:forward}, the induction is on the instructions,
for the backward simulation~\eqref{item:backward}, on the number of
iterations of the interleaving loop.

\paragraph{Initialization} An initial state of the simulation is:
$$%
\left(
  \state{%
    \ctxf
    {\simmain}
    {\env_{\simname}}
    {\cinit \app \{ \iwhile{\neg\terminated}{(\cselect\app\csim\app\cterminate)}\}}
  }
  {\heapsim}
\right)
$$
As we suppose (by construction) that initially, $\heapsimo = \heappar$
and that $\heapsims$ contains correctly allocated simulation blocks
for local variables and by the definition of a parallel initial state:
$(\state{\StacksP{\initname}}{\heappar^{\initname}})$ such that
$\forall t \in \setTid.~\stacks{\initname}{t} = \thdstacki$ the parts
\eqref{eq:equiv:heap} and \eqref{eq:equiv:vars} of the equivalence are
verified. The idea is then to show that the execution of $\cinit$
correctly establish \eqref{eq:equiv:pct:running},
\eqref{eq:equiv:pct:terminated}, \eqref{eq:equiv:wfstacks} and
\eqref{eq:equiv:simloop}.  In $\cinit$, we first move each program
counter to the first instruction of each main procedure, ensuring
\eqref{eq:equiv:pct:running} and \eqref{eq:equiv:pct:terminated} , and
then initialize the $\From$ address of each of these main procedures
to ensure \eqref{eq:equiv:wfstacks} (the base of the stack is
correctly modeled). Finally, we initialize \lstinline{terminated} to
false, since each thread must, at least, return from its main,
ensuring the \eqref{eq:equiv:simloop}. We have reached a state
$\stsim^0$ equivalent to
$\state{\StacksP{\initname}}{\heappar^{\initname}}$.

\begin{lemma}[Forward simulation on a single step]
  \label{lem:forward_1step}
  Let $\pprog$ be a safe parallel program and $\simprog$ its
  simulating program, $\stpar$ a parallel state that reaches $\stpar'$
  with an event $(t,\paction)$, $\stsim$ equivalent to $\stpar$, there
  exists a trace $\trname$ equivalent to $[(t,\paction)]$ that allows
  to reach $\stsim'$ equivalent to $\stpar'$.
\end{lemma}
\begin{sproof}
  By the equivalence relation, we know that $\stsim$ is of the form:
  $$
  \left(%
    \state%
    {\ctxf
      {\dots}
      {\env_{\simname}}
      {\iwhile{\neg\terminated}{(\cselect\app\csim\app\cterminate)}}
    }
    {\heapsim}
  \right)
  $$
  In the parallel semantics, we perform a step of reduction for the
  thread $t$, so its stack is not empty, and
  by~\eqref{eq:equiv:pct:running}, we know that
  $\heapsims[t](\pct) \neq 0$, and consequently, by
  \eqref{eq:equiv:simloop}, $\env_{\simname}(\terminated) =
  \vfalse$. We get the simulating program state:
  $$
  \left(
    \state%
    {\ctxf
      {\dots}
      {\env_{\simname}}
      {\cselect\app\csim\app\cterminate\app
        \{\iwhile{\neg\terminated}{(\cselect\app\csim\app\cterminate)}\}}
    }
    {\heapsim}
  \right)
  $$
  We then perform the reduction \selectRule. It generates an action
  $\acall{\Select}{[\loc_{\ptid};~\loc_{\pct}]}$ (the first action of
  $\trname$) that places $t$ at memory location $\ptid$, $t$ being an
  allowed choice for $\Select$ since $\heapsims[t](\pct) \neq 0$.

  At this step, we perform a case analysis depending on the executed
  instruction and prove that the execution reach a state where the
  parts \eqref{eq:equiv:heap} to \eqref{eq:equiv:wfstacks} of the
  equivalence are verified. Then, the execution of $\cterminate$
  updates the variable $\terminated$ by successively comparing the
  program counters to $0$. As we maintained
  \eqref{eq:equiv:pct:running} and \eqref{eq:equiv:pct:terminated}, we
  reach a state such that \eqref{eq:equiv:simloop} is verified:
  $$
  \left(%
    \state%
    {\ctxf
      {\dots}
      {\env_{\simname}}
      {\{\iwhile{\neg\terminated}{(\cselect\app\csim\app\cterminate)}\}}
    }
    {\heapsim''}
  \right)
  $$
  Moreover, actions generated during this loop are reads in
  $\heapsims$ and $\asilent$-actions (that are filtered). We reach,
  from $\stsim$ equivalent to $\stpar$, a state $\stsim'$ equivalent
  to $\stpar'$, with a trace $\trname$ equivalent to $[(t,\paction)]$.
\end{sproof}

\begin{lemma}[Backward simulation on a single step]
  \label{lem:backward_1step}
  Let $\simprog$ be the simulating program of a safe parallel program
  $\pprog$, $\stsim$ a sequential state that reach $\stsim'$ with a
  trace
  $\trname = (\acall{\textbf{select}}{[\loc_{\ptid};~\loc_{\pct}]}) ;
  \trname'$, such that $tr'$ does not contain call action to $select$,
  $\stpar$ equivalent to $\stsim$, there exists an action
  $(t,\paction)$ such that $[(t,\paction)]$ is equivalent to $\trname$
  that allows to reach $\stpar'$ equivalent to $\stsim'$.
\end{lemma}
\begin{sproof}Starting from $ \left(%
    \state%
    {\ctxf {\dots} {\env_{\simname}}
      {\{\iwhile{\neg\terminated}{(\cselect\app\csim\app\cterminate)}\}}
    } {\heapsim} \right) $, the simulation builds a trace
  $\trname=(\acall{\Select}{[\loc_{\ptid},~\loc_{\pct}]}) ; \trname'$
  so the condition is evaluated to $\vtrue$ (else we would not execute
  the loop, and the first action of the trace would not be
  realized). We also know that there exists $t$ such that
  $\heapsims(pct) \neq 0$. We know that in the original program, there
  exists $t$ such that $\stacks{}{t}$ is not empty, and we know the
  instruction $\instr$ that has to be performed by $t$.

  As $\pprog$ is safe, it does not block, the instruction $\instr$ of
  $t$ can be executed, so there exists a new parallel state $\stpar'$,
  reached with an action $(t,\paction)$. By
  lemma~\ref{lem:forward_1step}, we know that there exists a simulated
  state $\stname_{sim?}'$ equivalent to $\stpar'$ reached from
  $\stsim$ with a trace $tr_f$ equivalent to $[(t,\paction)]$. This
  trace $tr_f$ starts with an action
  $\acall{\textbf{select}}{[\loc_{\mathit{ptid}}]}$ equivalent to the
  one produced for $\trname$ and represents the execution of $\instr$
  by $t$, that is also simulated by our program $\simprog$. We can
  deduce that $\stname_{sim?}' = \stsim'$. As $\stpar'$ is equivalent
  to $\stname_{sim?}'$, it is also equivalent to $\stsim'$. Moreover
  $\trname = \trname_f$, $\trname_f$ is equivalent to
  $[(t,\paction)]$, so $\trname$ is equivalent to $[(t,\paction)]$.
\end{sproof}

The case analysis of each type of instruction is not presented since
it is a lot a details (that can be found in~\cite{BKL2017:RR}). The
ideas are quite the same from a proof to another: we show
that~\eqref{eq:equiv:vars} ensures we correctly replicate the local
variables in the simulating procedure, we deduce that the ``actual''
simulation instruction performs exactly the action that the original
one (which maintains \eqref{eq:equiv:heap} and \eqref{eq:equiv:vars}),
and finally we show that the program counter is correctly updated
according to the next instruction (\eqref{eq:equiv:pct:running}
and~\eqref{eq:equiv:pct:terminated}). The only tricky part is in
function call and returns where we have to ensure that $\From$ is
updated with respect to the original stack update and maintains
\eqref{eq:equiv:wfstacks}, for other instructions this part is trivial
since the stack does not change and $\From$ is not updated.


\section{Towards a Mechanized Proof of Correctness} 
\label{sec:coq}
We aim at mechanizing the proof of correctness using the proof
assistant \Coq. A first step to do so is to formalize both languages
and their semantics, as well as the transformation. The current state of the
development\footnote{Available at
  \url{http://frederic.loulergue.eu/ftp/cconc2seq-0.1alpha.tar.gz}}
includes this first step (about 3,000 lines of \Coq, one third being
proofs).

We have roughly 20\% devoted to supporting definitions and results
(about quite general data types and data structures used in the rest
of the formalization), 50\% to the syntax and semantics of the two
programming languages (about half of it comes from another project
with only slight modifications), the remaining 30\% focusing on the
formalization of the transformation and the statement of the
correctness theorem.

The syntax and semantics of the languages have a rather usual
formalization. As we seek reuse, we modeled the sequential semantics
so that it is parametrized by a set of ``external procedure''
definitions as it is found in some programming languages where the
signatures of some procedures are given but their implementation is
done in a foreign language. Here some procedures are not defined in
the programming language but are axiomatized by additional semantic
rules. {\tt select} is defined by such an external procedure
definition.

One important difference between the program definitions on paper and
in \Coq, is that in the mechanized version, all should be explicit. 
For example, in Section~\ref{sec:languages}, we leave implicit that
procedure names should not be duplicated in the list of procedure
definitions. 

The validity of procedure calls is used to define a relation on
procedures. For two procedures $p_1$ and $p_2$ of a program $\Prog$,
we have $p_1 \prec p_2$ if the body of $p_2$ contains a call to $p_1$.
To ensure that all procedures of a program are non-recursive, if is
sufficient to require that $\prec$ is well-founded. 
This property is necessary for two reasons. First, our simplified way to
simulate the call stack in the transformed code requires it. Second,
\Coq requires that all functions are {\em terminating}. It
automatically checks the termination of recursive functions when the
recursive calls are done on syntactical sub-terms of one of the
arguments of the function. In other cases, a proof of termination
should be given.



The next step of the mechanization is to define the equivalence
between states: the properties about the uniqueness of
procedure names and correct labelling are very important in this
regard. The final step will be to prove the correctness.


\section{Related Work} 
\label{sec:relatedwork}
Many model checking tools for concurrent programs are based on code
sequentialization. In~\cite{QW2004:PLDI}, Qadeer and Wu present, for
the C language, a transformation from parallel to sequential code that
allows the use of existing model checkers for sequential systems. This
bounded model checking has been generalized to any context bounds with
CSeq~\cite{LR2008:CAV} and dynamic thread
creation~\cite{FIP2013:ASE}. While bounded, such an approach is still
efficient to find bugs in concurrent programs. Regarding code
transformation, these approaches differ from ours since in each
thread, functions are inlined in the main function, loops are unrolled
and $k$ copies of the global memory are kept for a bound of $K$ thread
context switching.  To avoid creating these copies, allow dynamic
memory allocation and improve performances, Fisher et
al.~\cite{ITF2014:CAV} propose a lazy version of these tools called
LazySeq that shows high performances on known benchmarks. Other
authors choose to bound memory accesses instead of context
switching~\cite{TIF2015:TACAS}.

While efficient to find bugs, these approaches are not suited to prove
safety, which is the main reason we aimed at supporting the \WP plugin of
\framac.  In~\cite{NFS2016:ATVA}, Nguyen et al. further generalize
LazySeq to unbounded concurrent programs allowing safety checking. The
approach for code generation is somehow dual to ours: instead of
splitting original functions into smaller functions for each statement
and adding the context switching management in an interleaving loop,
context switching is modeled inside each function to obtain a behavior
where each call to the function will execute a step of execution and
then return (and where local variables become static).  All these
approaches consider a sequentially consistent memory model, as we do,
while other work is aimed at supporting weaker behaviors as well, 
e.g.~\cite{TNI2016:FMCAD}.

Why3 is a deductive verification tool that proposes Why-ML, a language
for writing programs and assertions, and a verification condition
generator as well as translations of these conditions as input to a
wide variety of automated provers.  Fortin and Gava~\cite{FOR2013:PHD}
used program transformation to perform deductive proof of bulk
synchronous parallel (BSP) programs. In this work, the original
annotated parallel program is compiled into an equivalent sequential
Why-ML program. The deductive proof is then performed using the
original Why-ML VCGen, designed for sequential programs. The
transformation is written and proved using the \Coq proof
assistant. If the software context is very close to our proposal, the
parallelism models are very different and thus so are the
transformations: A BSP program is a sequence of super-steps, and the
parallelism occurs inside each super-step. Inside a super-step each
thread computes using only the data it holds in memory, then
communicates with other threads, but the results of these
communications (by message passing) are not effective before the end
of the super-step (that contains a synchronization barrier).

The way we transform code and specification makes the use of WP after
the transformation closely related to Owicki-Gries
method~\cite{OG1976:CACM}.  Indeed, for each instruction, we have to
ensure that it is compatible with any state of the global system that
can be reached at some program point.  This property is modeled by a
global invariant. Unlike~\cite{OG1976:CACM}, this compatibility is not
verified by visiting the proof tree. Owicki-Gries method has been
formalized in Isabelle/HOL~\cite{NP1999:FASE} and one of its variants
has been used for verification of operating
systems~\cite{ALM2016:ITP}. So, even if it can generate a lot of
verification conditions, it is still usable in practice for real-life
code.





\section{Conclusion} 
\label{sec:conclusion}
The contribution of this paper is the correctness proof of the
principle of a code transformation used to verify concurrent C code
through a sequential C program that simulates it, in the context of a
sequentially consistent memory model. This proof is done under the
assumption that the source program does not allocate memory and does
not contain any recursive call.  This proof has three main concerns:
the heap of the source program should be correctly replicated in the
transformed program; the local environments of the source
program should be correctly simulated by the {\em global heap} of the
transformed program; the execution context of the source program should be
correctly modelled by the memory location that stores a kind of
program counter and the memory locations that model a simplified call
stack.

The proof relies on the fact that in a way the simulating code mimics
the operational semantics of the concurrent program with its own
sequential instructions, but in a simplified version (in particular
because we do not really need to simulate a call stack). Moreover all
the simulating code is deterministic but the code that simulates
thread switching.  We aim at the mechanization of this proof in the
interactive theorem prover \Coq. A non-trivial first step was to
formalize the languages and their semantics, as well as the
transformation.  The next step will be to write the correctness proof
itself with \Coq.

The \toolname plugin does not only transform the code to verify.
\toolname provides extensions to the \ACSL behavioral specification
language in order to write contracts for concurrent C programs. These
assertions are also transformed by the plugin. Ultimately we would
like to formalize axiomatic semantics for the parallel and sequential
languages, and verify that the transformation of both code and
assertions is such that a proof (using the sequential axiomatic
semantics) of a simulating program allows to build a proof (using the
parallel axiomatic semantics) of the source concurrent program.  This
is however a long term goal.  Future work also includes extensions to
the plugin itself that could also be verified as extensions of the
current formal framework.


\bibliographystyle{eptcs} 
\bibliography{bibliography}

\begin{thebibliography}{10}
\providecommand{\bibitemdeclare}[2]{}
\providecommand{\surnamestart}{}
\providecommand{\surnameend}{}
\providecommand{\urlprefix}{Available at }
\providecommand{\url}[1]{\texttt{#1}}
\providecommand{\href}[2]{\texttt{#2}}
\providecommand{\urlalt}[2]{\href{#1}{#2}}
\providecommand{\bibinfo}[2]{#2}

\bibitemdeclare{inproceedings}{ALM2016:ITP}
\bibitem{ALM2016:ITP}
\bibinfo{author}{June \surnamestart Andronick\surnameend},
  \bibinfo{author}{Corey \surnamestart Lewis\surnameend},
  \bibinfo{author}{Daniel \surnamestart Matichuk\surnameend},
  \bibinfo{author}{Carroll \surnamestart Morgan\surnameend} \&
  \bibinfo{author}{Christine \surnamestart Rizkallah\surnameend}
  (\bibinfo{year}{2016}): \emph{\bibinfo{title}{Proof of {OS} Scheduling
  Behavior in the Presence of Interrupt-Induced Concurrency}}.
\newblock In: {\sl \bibinfo{booktitle}{Interactive Theorem Proving - 7th
  International Conference, {ITP} 2016, Nancy, France, August 22-25, 2016,
  Proceedings}}, pp. \bibinfo{pages}{52--68},
  \doi{10.1007/978-3-319-43144-4_4}.

\bibitemdeclare{manual}{ACSL:20151002}
\bibitem{ACSL:20151002}
\bibinfo{author}{Patrick \surnamestart Baudin\surnameend},
  \bibinfo{author}{Jean~C. \surnamestart Filli\^{a}tre\surnameend},
  \bibinfo{author}{Pascal \surnamestart Cuoq\surnameend},
  \bibinfo{author}{Claude \surnamestart March\'{e}\surnameend},
  \bibinfo{author}{Benjamin \surnamestart Monate\surnameend},
  \bibinfo{author}{Yannick \surnamestart Moy\surnameend} \&
  \bibinfo{author}{Virgile \surnamestart Prevosto\surnameend}
  (\bibinfo{year}{2015}): \emph{\bibinfo{title}{{ACSL}: {ANSI/ISO} C
  Specification Language}}.
\newblock \bibinfo{note}{\url{http://frama-c.com/download.html}}.

\bibitemdeclare{inproceedings}{BKL2015:FMICS}
\bibitem{BKL2015:FMICS}
\bibinfo{author}{Allan \surnamestart Blanchard\surnameend},
  \bibinfo{author}{Nikolai \surnamestart Kosmatov\surnameend},
  \bibinfo{author}{Matthieu \surnamestart Lemerre\surnameend} \&
  \bibinfo{author}{Fr\'{e}d\'{e}ric \surnamestart Loulergue\surnameend}
  (\bibinfo{year}{2015}): \emph{\bibinfo{title}{A Case Study on Formal
  Verification of the {A}naxagoros Hypervisor Paging System with {F}rama-{C}}}.
\newblock In: {\sl \bibinfo{booktitle}{International Workshop on Formal Methods
  for Industrial Critical Systems (FMICS)}}, \bibinfo{series}{LNCS},
  \bibinfo{publisher}{Springer}, \bibinfo{address}{Oslo, Norway}, pp.
  \bibinfo{pages}{15--30}, \doi{10.1007/978-3-319-19458-5_2}.

\bibitemdeclare{inproceedings}{BKL2016:SCAM}
\bibitem{BKL2016:SCAM}
\bibinfo{author}{Allan \surnamestart Blanchard\surnameend},
  \bibinfo{author}{Nikolai \surnamestart Kosmatov\surnameend},
  \bibinfo{author}{Matthieu \surnamestart Lemerre\surnameend} \&
  \bibinfo{author}{Fr\'{e}d\'{e}ric \surnamestart Loulergue\surnameend}
  (\bibinfo{year}{2016}): \emph{\bibinfo{title}{Conc2Seq: A {F}rama-{C} Plugin
  for Verification of Parallel Compositions of {C} Programs}}.
\newblock In: {\sl \bibinfo{booktitle}{16th IEEE International Working
  Conference on Source Code Analysis and Manipulation (SCAM)}},
  \bibinfo{publisher}{IEEE}, \bibinfo{address}{Raleigh, NC, USA}, pp.
  \bibinfo{pages}{67--72}, \doi{10.1109/SCAM.2016.18}.

\bibitemdeclare{techreport}{BKL2017:RR}
\bibitem{BKL2017:RR}
\bibinfo{author}{Allan \surnamestart Blanchard\surnameend},
  \bibinfo{author}{Nikolai \surnamestart Kosmatov\surnameend} \&
  \bibinfo{author}{Fr\'{e}d\'{e}ric \surnamestart Loulergue\surnameend}
  (\bibinfo{year}{2017}): \emph{\bibinfo{title}{{Concurrent Program
  Verification by Code Transformation: Correctness}}}.
\newblock \bibinfo{type}{Research Report} \bibinfo{number}{RR-2017-03},
  \bibinfo{institution}{LIFO, Universit\'{e} d'Orl\'{e}ans}.
\newblock
  \urlprefix\url{http://www.univ-orleans.fr/lifo/prodsci/rapports/RR/RR2017/RR-2017-03.pdf}.

\bibitemdeclare{inproceedings}{FIP2013:ASE}
\bibitem{FIP2013:ASE}
\bibinfo{author}{Bernd \surnamestart Fischer\surnameend}, \bibinfo{author}{Omar
  \surnamestart Inverso\surnameend} \& \bibinfo{author}{Gennaro \surnamestart
  Parlato\surnameend} (\bibinfo{year}{2013}): \emph{\bibinfo{title}{CSeq: {A}
  concurrency pre-processor for sequential {C} verification tools}}.
\newblock In: {\sl \bibinfo{booktitle}{2013 28th {IEEE/ACM} International
  Conference on Automated Software Engineering, {ASE} 2013, Silicon Valley, CA,
  USA, November 11-15, 2013}}, pp. \bibinfo{pages}{710--713},
  \doi{10.1109/ASE.2013.6693139}.

\bibitemdeclare{phdthesis}{FOR2013:PHD}
\bibitem{FOR2013:PHD}
\bibinfo{author}{Jean \surnamestart Fortin\surnameend} (\bibinfo{year}{2013}):
  \emph{\bibinfo{title}{{BSP-Why: a Tool for Deductive Verification of BSP
  Programs}}}.
\newblock Ph.D. thesis, \bibinfo{school}{Universit\'{e} Paris-Est Cr\'{e}teil,
  LACL}.
\newblock \urlprefix\url{http://hal.archives-ouvertes.fr/tel-00974977/}.

\bibitemdeclare{inproceedings}{ITF2014:CAV}
\bibitem{ITF2014:CAV}
\bibinfo{author}{Omar \surnamestart Inverso\surnameend},
  \bibinfo{author}{Ermenegildo \surnamestart Tomasco\surnameend},
  \bibinfo{author}{Bernd \surnamestart Fischer\surnameend},
  \bibinfo{author}{Salvatore \surnamestart {La Torre}\surnameend} \&
  \bibinfo{author}{Gennaro \surnamestart Parlato\surnameend}
  (\bibinfo{year}{2014}): \emph{\bibinfo{title}{Bounded Model Checking of
  Multi-threaded {C} Programs via Lazy Sequentialization}}.
\newblock In: {\sl \bibinfo{booktitle}{Computer Aided Verification - 26th
  International Conference, {CAV} 2014, Held as Part of the Vienna Summer of
  Logic, {VSL} 2014, Vienna, Austria, July 18-22, 2014. Proceedings}}, pp.
  \bibinfo{pages}{585--602}, \doi{10.1007/978-3-319-08867-9_39}.

\bibitemdeclare{article}{KKP2015:FAC}
\bibitem{KKP2015:FAC}
\bibinfo{author}{Florent \surnamestart Kirchner\surnameend},
  \bibinfo{author}{Nikolai \surnamestart Kosmatov\surnameend},
  \bibinfo{author}{Virgile \surnamestart Prevosto\surnameend},
  \bibinfo{author}{Julien \surnamestart Signoles\surnameend} \&
  \bibinfo{author}{Boris \surnamestart Yakobowski\surnameend}
  (\bibinfo{year}{2015}): \emph{\bibinfo{title}{{F}rama-{C}: {A} software
  analysis perspective}}.
\newblock {\sl \bibinfo{journal}{Formal Asp. Comput.}}
  \bibinfo{volume}{27}(\bibinfo{number}{3}), pp. \bibinfo{pages}{573--609},
  \doi{10.1007/s00165-014-0326-7}.

\bibitemdeclare{inproceedings}{KPS2013:TAP}
\bibitem{KPS2013:TAP}
\bibinfo{author}{Nikolai \surnamestart Kosmatov\surnameend},
  \bibinfo{author}{Virgile \surnamestart Prevosto\surnameend} \&
  \bibinfo{author}{Julien \surnamestart Signoles\surnameend}
  (\bibinfo{year}{2013}): \emph{\bibinfo{title}{A Lesson on Proof of Programs
  with {Frama-C}. {I}nvited Tutorial Paper}}.
\newblock In: {\sl \bibinfo{booktitle}{TAP}}, {\sl \bibinfo{series}{LNCS}}
  \bibinfo{volume}{7942}, \bibinfo{publisher}{Springer}, pp.
  \bibinfo{pages}{168--177}, \doi{10.1007/978-3-642-38916-0_10}.

\bibitemdeclare{inproceedings}{LR2008:CAV}
\bibitem{LR2008:CAV}
\bibinfo{author}{Akash \surnamestart Lal\surnameend} \&
  \bibinfo{author}{Thomas~W. \surnamestart Reps\surnameend}
  (\bibinfo{year}{2008}): \emph{\bibinfo{title}{Reducing Concurrent Analysis
  Under a Context Bound to Sequential Analysis}}.
\newblock In: {\sl \bibinfo{booktitle}{Computer Aided Verification, 20th
  International Conference, {CAV} 2008, Princeton, NJ, USA, July 7-14, 2008,
  Proceedings}}, pp. \bibinfo{pages}{37--51},
  \doi{10.1007/978-3-540-70545-1_7}.

\bibitemdeclare{article}{L79:IEEE}
\bibitem{L79:IEEE}
\bibinfo{author}{L.~\surnamestart Lamport\surnameend} (\bibinfo{year}{1979}):
  \emph{\bibinfo{title}{{How to Make a Multiprocessor Computer That Correctly
  Executes Multiprocess Program}}}.
\newblock {\sl \bibinfo{journal}{IEEE Trans. Comput.}}
  \bibinfo{volume}{28}(\bibinfo{number}{9}), pp. \bibinfo{pages}{690--691},
  \doi{10.1109/TC.1979.1675439}.

\bibitemdeclare{article}{LER2009:JAR}
\bibitem{LER2009:JAR}
\bibinfo{author}{Xavier \surnamestart Leroy\surnameend} (\bibinfo{year}{2009}):
  \emph{\bibinfo{title}{A Formally Verified Compiler Back-end}}.
\newblock {\sl \bibinfo{journal}{Journal of Automated Reasoning}}
  \bibinfo{volume}{43}(\bibinfo{number}{4}), pp. \bibinfo{pages}{363--446},
  \doi{10.1007/s10817-009-9155-4}.

\bibitemdeclare{inproceedings}{NFS2016:ATVA}
\bibitem{NFS2016:ATVA}
\bibinfo{author}{Truc~L. \surnamestart Nguyen\surnameend},
  \bibinfo{author}{Bernd \surnamestart Fischer\surnameend},
  \bibinfo{author}{Salvatore \surnamestart {La Torre}\surnameend} \&
  \bibinfo{author}{Gennaro \surnamestart Parlato\surnameend}
  (\bibinfo{year}{2016}): \emph{\bibinfo{title}{Lazy Sequentialization for the
  Safety Verification of Unbounded Concurrent Programs}}.
\newblock In: {\sl \bibinfo{booktitle}{Automated Technology for Verification
  and Analysis - 14th International Symposium, {ATVA} 2016, Chiba, Japan,
  October 17-20, 2016, Proceedings}}, pp. \bibinfo{pages}{174--191},
  \doi{10.1007/978-3-319-46520-3_12}.

\bibitemdeclare{inproceedings}{NP1999:FASE}
\bibitem{NP1999:FASE}
\bibinfo{author}{T.~\surnamestart Nipkow\surnameend} \&
  \bibinfo{author}{L.~\surnamestart Prensa~Nieto\surnameend}
  (\bibinfo{year}{1999}): \emph{\bibinfo{title}{{Owicki/Gries in
  Isabelle/HOL}}}.
\newblock In \bibinfo{editor}{J.-P. \surnamestart Finance\surnameend}, editor:
  {\sl \bibinfo{booktitle}{Fundamental Approaches to Software Engineering,
  Second International Conference (FASE'99)}}, \bibinfo{volume}{LNCS 1577},
  \bibinfo{publisher}{Springer}, pp. \bibinfo{pages}{188--203},
  \doi{10.1007/978-3-540-49020-3_13}.

\bibitemdeclare{article}{OG1976:CACM}
\bibitem{OG1976:CACM}
\bibinfo{author}{S.~\surnamestart Owicki\surnameend} \&
  \bibinfo{author}{D.~\surnamestart Gries\surnameend} (\bibinfo{year}{1976}):
  \emph{\bibinfo{title}{Verifying properties of parallel programs: an axiomatic
  approach}}.
\newblock {\sl \bibinfo{journal}{Communications of the ACM}}
  \bibinfo{volume}{19}(\bibinfo{number}{5}), pp. \bibinfo{pages}{279--285},
  \doi{10.1145/360051.360224}.

\bibitemdeclare{inproceedings}{QW2004:PLDI}
\bibitem{QW2004:PLDI}
\bibinfo{author}{Shaz \surnamestart Qadeer\surnameend} \&
  \bibinfo{author}{Dinghao \surnamestart Wu\surnameend} (\bibinfo{year}{2004}):
  \emph{\bibinfo{title}{{KISS:} keep it simple and sequential}}.
\newblock In: {\sl \bibinfo{booktitle}{Proceedings of the {ACM} {SIGPLAN} 2004
  Conference on Programming Language Design and Implementation 2004,
  Washington, DC, USA, June 9-11, 2004}}, pp. \bibinfo{pages}{14--24},
  \doi{10.1145/996841.996845}.

\bibitemdeclare{inproceedings}{TIF2015:TACAS}
\bibitem{TIF2015:TACAS}
\bibinfo{author}{Ermenegildo \surnamestart Tomasco\surnameend},
  \bibinfo{author}{Omar \surnamestart Inverso\surnameend},
  \bibinfo{author}{Bernd \surnamestart Fischer\surnameend},
  \bibinfo{author}{Salvatore \surnamestart {La Torre}\surnameend} \&
  \bibinfo{author}{Gennaro \surnamestart Parlato\surnameend}
  (\bibinfo{year}{2015}): \emph{\bibinfo{title}{Verifying Concurrent Programs
  by Memory Unwinding}}.
\newblock In: {\sl \bibinfo{booktitle}{Tools and Algorithms for the
  Construction and Analysis of Systems - 21st International Conference, {TACAS}
  2015, Held as Part of the European Joint Conferences on Theory and Practice
  of Software, {ETAPS} 2015, London, UK, April 11-18, 2015. Proceedings}}, pp.
  \bibinfo{pages}{551--565}, \doi{10.1007/978-3-662-46681-0_52}.

\bibitemdeclare{inproceedings}{TNI2016:FMCAD}
\bibitem{TNI2016:FMCAD}
\bibinfo{author}{Ermenegildo \surnamestart Tomasco\surnameend},
  \bibinfo{author}{Truc~Nguyen \surnamestart Lam\surnameend},
  \bibinfo{author}{Omar \surnamestart Inverso\surnameend},
  \bibinfo{author}{Bernd \surnamestart Fischer\surnameend},
  \bibinfo{author}{Salvatore~La \surnamestart Torre\surnameend} \&
  \bibinfo{author}{Gennaro \surnamestart Parlato\surnameend}
  (\bibinfo{year}{2016}): \emph{\bibinfo{title}{Lazy Sequentialization for TSO
  and PSO via Shared Memory Abstractions}}.
  \newblock In: {\sl \bibinfo{booktitle}{Formal Methods in Computer-Aided Design
      (FMCAD)}},
  \doi{10.1109/FMCAD.2016.7886679}.

\end{thebibliography}


\end{document}